\documentclass[fleqn,usenatbib]{mnras}

\usepackage{newtxtext,newtxmath}
\usepackage[T1]{fontenc}

\DeclareRobustCommand{\VAN}[3]{#2}
\let\VANthebibliography\thebibliography
\def\thebibliography{\DeclareRobustCommand{\VAN}[3]{##3}\VANthebibliography}

\usepackage{graphicx}	% Including figure files
\usepackage{amsmath}	% Advanced maths commands
\usepackage{breakurl}
\usepackage{color}
\usepackage{pdflscape}
\usepackage{multirow}
\usepackage{mathtools}  % for multilines in eq.1
\usepackage{longtable}
\usepackage{siunitx}    % for alignment by \pm in table A2
\newcommand{\g}{$\gamma$}
\newcommand{\dg}{^{\circ}}
\newcommand{\NMEAS}{5068}
\newcommand{\NSRC}{222}
\DeclareMathOperator\erf{erf}
\renewcommand{\theenumi}{\arabic{enumi}}
\title[RoboPol: monitoring data]{RoboPol: AGN polarimetric monitoring data}

\author[D. Blinov et al.]
{D. Blinov$^{1,2,3}$\thanks{E-mail: blinov@ia.forth.gr}, S. Kiehlmann$^{1,2}$, V. Pavlidou$^{1,2}$, G. V. Panopoulou$^{4}$\thanks{Hubble Fellow}, R. Skalidis$^{1,2}$, 
\newauthor
E. Angelakis$^{5}$, C. Casadio$^{1,2,6}$, E. N. Einoder$^{4}$, T. Hovatta$^{7,8}$, K. Kokolakis$^{9,2}$,
\newauthor
A. Kougentakis$^{1}$, A. Kus$^{10}$, N. Kylafis$^{2,1}$, E. Kyritsis$^{2,1}$, A. Lalakos$^{11}$, I. Liodakis$^{7}$,
\newauthor
S. Maharana$^{12}$, E. Makrydopoulou$^{2}$, N. Mandarakas$^{1,2}$, G. M. Maragkakis$^{13,2}$, I. Myserlis$^{6}$,
\newauthor
I. Papadakis$^{1,2}$, G. Paterakis$^{1}$, T. J. Pearson$^{4}$, A. N. Ramaprakash$^{12,1,4}$, A.\,C.\,S. Readhead$^{4}$,
\newauthor
P. Reig$^{2,1}$, A. S\l{}owikowska$^{10}$, K. Tassis$^{1,2}$, K. Xexakis$^{2}$, M. {\. Z}ejmo$^{14}$, J.~A.~Zensus$^{6}$\\
$^{1}$Institute of Astrophysics, Foundation for Research and Technology-Hellas, Voutes, 71110 Heraklion, Greece\\
$^{2}$Department of Physics, 
University of Crete, 71003, Heraklion, Greece\\
$^{3}$Astronomical Institute, St. Petersburg State University, Universitetsky pr. 28, Petrodvoretz, 
198504 St. Petersburg, Russia \\
$^{4}$Cahill Center for Astronomy and Astrophysics, California Institute of Technology, 1200 E California Blvd, MC 350-17,\\Pasadena CA, 91125, USA\\
$^{5}$ Section of Astrophysics, Astronomy \& Mechanics, Department of Physics, National and Kapodistrian University of Athens,\\ Panepistimiopolis Zografos 15784, Greece\\
$^{6}$Max-Planck-Institut f\"{u}r Radioastronomie, Auf dem H\"{u}gel 69, 53121 Bonn, Germany\\
$^{7}$Finnish Centre for Astronomy with ESO, FINCA, University of Turku, Quantum, Vesilinnantie 5, FI-20014, Finland\\
$^{8}$Aalto University Mets\"ahovi Radio Observatory, Mets\"ahovintie 114, FI-02540 Kylm\"al\"a, Finland\\
$^{9}$Geodesy and Geomatics Engineering Laboratory, Technical University of Crete, GR-73100 Chania, Greece\\
$^{10}$Institute of Astronomy, Faculty of Physics, Astronomy and Informatics, Nicolaus Copernicus University in Toru\'n, Grudziadzka 5,\\PL-87-100 Toru\'n, Poland\\
$^{11}$Center for Interdisciplinary Exploration \& Research in Astrophysics (CIERA), Northwestern University, Evanston, IL 602, USA\\
$^{12}$Inter-University Centre for Astronomy and Astrophysics, Post Bag 4, Ganeshkhind, Pune - 411 007, India\\
$^{13}$Institute of Electronic
Structure and Laser, Foundation for Research and Technology-Hellas, Voutes, 71110 Heraklion, Greece\\
$^{14}$Janusz Gil Institute of Astronomy, University of Zielona G\'{o}ra, Prof. Szafrana 2, PL-65-516 Zielona G\'{o}ra, Poland\\
}

\date{Accepted XXX. Received YYY; in original form ZZZ}

\pubyear{2016}

\begin{document}
\label{firstpage}
\pagerange{\pageref{firstpage}--\pageref{lastpage}}
\maketitle

\begin{abstract}
We present uniformly reprocessed and re-calibrated data from the {\em RoboPol} programme of optopolarimetric monitoring of active galactic nuclei (AGN), covering observations between 2013, when the instrument was commissioned, and 2017. In total, the dataset presented in this paper includes $\NMEAS$ observations of $\NSRC$ AGN with Dec > $-$25\degr. We describe the current version of the RoboPol pipeline that was used to process and calibrate the entire dataset, and we make the data publicly available for use by the astronomical community. Average quantities summarising optopolarimetric behaviour  (average degree of polarization, polarization variability index) are also provided for each source we have observed and for the time interval we have followed it. 
\end{abstract}


%%%%%%%%%%%%%%%%%%%%%%%%%%%%%%%%%%%%%%%%%%%%%%%%%%

%%%%%%%%%%%%%%%%% BODY OF PAPER %%%%%%%%%%%%%%%%%%

\section{Introduction} \label{sec:introduction}

The {\em RoboPol} Collaboration\footnote{\url{http://robopol.org}} monitored the optical linear polarization and brightness of a large sample of active galactic nuclei (AGN) from 2013 to 2017, using the {\em RoboPol} polarimeter \citep{instrumentpaper}, which was developed for this project, and which is installed at the 1.3m telescope of the Skinakas Observatory in Crete, Greece. The main science goal of the {\em RoboPol} project was to explore a possible link between optical polarization behaviour, particularly the electric vector position angle (EVPA) rotations, and flares in the \g-ray emission of blazars. The main project was run between 2013 May and 2015 November. During this period the monitoring was focused on two well-defined samples: the main sample, consisting of \g-ray loud AGN detected in the {\em Fermi}--LAT Second Source Catalogue \citep[2FGL, ][]{Nolan2012}; and the control sample, consisting of otherwise similar AGN, which however had not been detected by {\em Fermi}--LAT. The sample selection is discussed in \S\ref{sec:sample} and in greater detail in  \cite{Pavlidou2014}. In addition to the main monitoring programme, we observed additional sources for other projects, also described in \S\ref{sec:sample}. Major results based on the data set presented in this paper were published in  \cite{Pavlidou2014,Blinov2015,Blinov2016a,Blinov2016b,Blinov2018,Angelakis2016,Hovatta2016,Kiehlmann2017,Liodakis2017}.

In this paper, we present a complete and uniform reprocessing of all observations in the 5-year AGN monitoring dataset, using the latest version of the {\em RoboPol} pipeline. The samples of sources that have been included in our monitoring during the programme are summarised in \S \ref{sec:sample}. Information on the {\em RoboPol} polarimeter and the telescope, where it is mounted, is given in \S \ref{sec:telesc}. The current version of the {\em RoboPol} pipeline that has been used to process all data presented in this paper is described in \S \ref{sec:pipeline}. Details on the standard stars used for calibration are given in \S \ref{sec:polar}. The reader is alerted to caveats that have arisen in the processing of specific sources in \S \ref{sec:ind_src}. The data we release for each source, including individual observations and summary statistics quantifying the optopolarimetric behaviour of each source are described in \S \ref{sec:data}. This data are made publicly available for use by the astronomical community. Our data policy is discussed in \S \ref{sec:datapolicy}. 
The values of EVPA in this work are measured from North to East following the IAU convention \citep{EVPAconv}. All monitoring data made available in this paper are not corrected for positive bias of the polarization fraction \citep{Serkowski1958}. 

With this paper we publish $\NMEAS$ polarimetric measurements of $\NSRC$ AGN located at Dec >$-$25\degr obtained during the 2013 -- 2017 observing seasons.

\section{Observing samples} \label{sec:sample}

We provide data for sources belonging to several observing samples as indicated by the 'sample' flag in the data Table \ref{ap:a}. Here, we briefly describe these samples, and we refer the reader to the corresponding results-and-analysis papers for more details. 

During, and immediately after, the {\em RoboPol} polarimeter commissioning, we performed a month-long single-epoch survey of two unbiased samples of 89 \g-ray--loud (Sample Identifier SID=1, see Table \ref{ap:a}) and 15 \g-ray--quiet (SID=2) sources. The details of the sample selection are described in \cite{Pavlidou2014}. 

During the main monitoring programme, we regularly observed sources belonging to two samples: the 'main' \g-ray--loud sample (SID=3), consisting of 62 sources selected by placing a photon-flux cut on sources from the Second Fermi LAT Catalogue \citep[2FGL,][]{Nolan2012} and an additional optical magnitude cut, as well as constraints on source visibility and separation from other field sources; and the 'control' \g-ray--quiet sample (SID=4), of 15 sources selected based on their radio variability properties and absence from 2FGL. Control sample sources were selected from non-2FGL sources from CGRaBS \citep[Candidate Gamma-Ray Blazar Survey,][]{Healey2008} placing a 15 GHz flux density cut, and identical optical magnitude, visibility, and field-quality cuts as for the \g-ray--loud sample. Among all sources satisfying these criteria, we selected the ones that were most variable in radio, as quantified by their 15~GHz modulation index  \citep{Richards2011}. 
Slight changes to the initial 'control' sample were made in \cite{Blinov2016a,Blinov2016b,Blinov2018} because two of the \g-ray--quiet sources appeared in the Third Fermi LAT Catalogue \citep[3FGL][]{Acero2015}. For this reason we added two more sources satisfying our selection criteria that were not present in any of the Fermi LAT catalogues. 
  
In addition to the 'main' and 'control' samples, during 2013 -- 2014 we monitored a sample of 24 individually-selected sources of high interest. Moreover, 44 other AGN were observed for different projects along the project execution. These sources that have {\em not} been selected with uniform criteria are assigned SID=5. In 2014 we also conducted a distinct polarization monitoring programme focused on intermediate- and high-synchrotron peaked BL Lac objects. This programme included 29 TeV-detected sources and 19 non-TeV sources, collectively described in Table \ref{ap:a} with SID=6. Details of the selection criteria for these samples are described in \cite{Hovatta2016}.

\section{Telescope and polarimeter} \label{sec:telesc}
\begin{figure}
 \centering
 \includegraphics[width=0.38\textwidth]{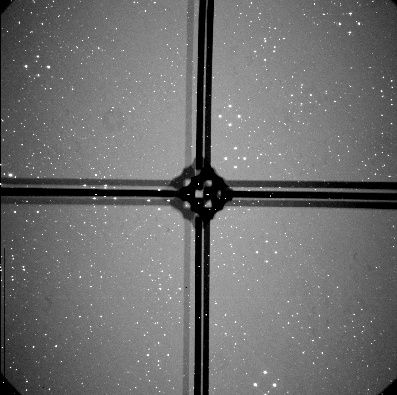}
 \caption{An example of a {\em RoboPol} image. Each point in the sky is mapped to four spots on the CCD. A focal plane mask, held in place by four support legs, reduces the sky background level for the central target.}
 \label{fig:shot}
\end{figure}

Optical polarimetric observations were performed using the 1.3-m telescope at the Skinakas Observatory\footnote{\url{http://skinakas.physics.uoc.gr}} in Crete (1750 m.a.s.l., 35\degr 12\arcmin 43\arcsec N, 24\degr 53\arcmin 57\arcsec E). The telescope is equipped with the {\em RoboPol} imaging polarimeter \citep{instrumentpaper}, which was designed specifically for this monitoring programme. {\em RoboPol} is comprised of two adjacent half-wave retarders with fast-axes rotated by $67.5\dg$ with respect to each other. They 
are followed by two Wollaston prisms with orthogonal fast-axes. This configuration splits every incident ray into two pairs of spots on the CCD that carry information about Q/I and U/I normalised Stokes parameters in the instrument reference frame \cite[see Eq.~1 in ][]{King2014}. In the case when a photometric standard with known magnitude is in the field, the total intensity or Stokes I parameter can be obtained together with Q/I and U/I with a single exposure. Unfortunately, only a small fraction of sources in our samples has accurately measured robust photometric standards in their fields. Therefore, here we present only normalised Stokes parameters without the total flux data. Since the instrument has no moving parts other than the filter wheel, it is relieved of random and systematic errors due to sky changes between measurements, imperfect alignment and non-uniformity of rotating optical elements. In order to increase the SNR (signal-to-noise ratio) for the central target 
measurements, a mask of a special shape was introduced in the centre of the focal plane. An example of an image obtained with {\em RoboPol} is shown in Fig.~\ref{fig:shot}.

The polarimeter is equipped with a $2048\times2048$ pixels ANDOR DW436 CCD with a 13.5 $\mu$m pixel size. It provides a scale of 0.435 arcsec pixel$^{-1}$ and a field of view (FoV) of $13 \times 13 $ arcmin$^{2}$. All observations described in this paper were made with
a Johnson-Cousins {\it R}-band filter.

\section{RoboPol pipeline} \label{sec:pipeline}
The {\em RoboPol} pipeline was planned and implemented to be capable of working with no user intervention.  It is able to derive the magnitude and linear polarization of every unobscured source in the field (i.e. every source that is not blocked by the focal plane mask and its holders). The pipeline is written in Python, with some procedures written in Cython to improve performance. A detailed description of the pipeline is presented by \cite{King2014}. Upgrades in the pipeline designed to optimize performance for field sources are discussed in \cite{Panopoulou2015}. 
In the following sections we describe only recent upgrades in the pipeline that have not been described elsewhere.

\subsection{Upgrades of the pipeline} \label{subsec:upgr}
\subsubsection{Mask location and background estimates} \label{subsubsec:locmask}

\begin{figure}
 \centering
 \includegraphics[width=0.36\textwidth]{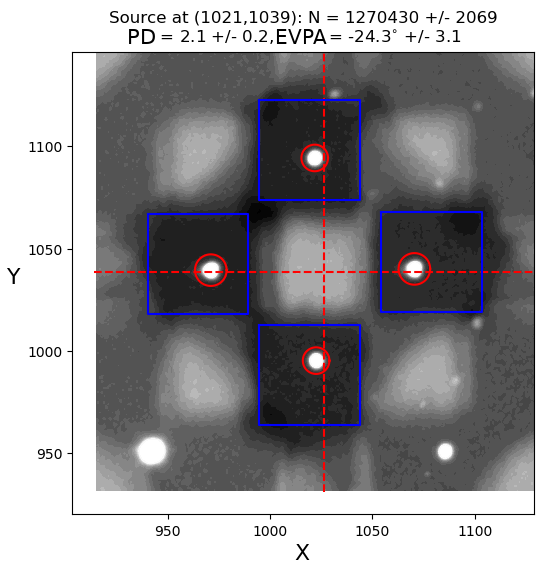}
 \caption{Masked central area of the RoboPol image with the central science target. The red circles are the photometry apertures. The blue squares are the automatically identified background estimation areas for
the central science target.}
 \label{fig:rbpl_src}
\end{figure}

The main science targets are positioned within the central, masked area of the {\em RoboPol} image shown in Fig.~\ref{fig:rbpl_src}. The sky background for each source spot is estimated within the corresponding masked areas (the blue squares in Fig.~\ref{fig:rbpl_src}). The four squares pattern is automatically identified by the pipeline. First versions of the pipeline considered this pattern as a whole. However, it was found that the relative positions of the shadows are variable within $\sim 10$~pix. Therefore, the latest version of the pipeline locates each square separately. This provides more accurate histograms of the background counts, preventing biasing of the background estimate towards brighter values.

The default background estimation method used in the pipeline has been changed to a procedure similar to the one used in SExtractor \citep{Bertin1996}. The procedure consists of  the following steps.  (1)  The histogram of pixel values within the masked square is iteratively sigma-clipped, with $2.7\sigma$ as the limit. (2) If the sigma-clipped $\sigma$ changed more than 20 per cent from its initial value, the histogram is considered to be skewed, and the background value is estimated using  ${\rm Mode} = 2.5\times {\rm Median} - 1.5 \times {\rm Mean}$. Otherwise, the mean of the clipped histogram is taken as the value for the background. 

The background histograms and calculated background values for all measurements in this paper have been visually inspected. In the cases where the default method provided unsatisfactory results, the images were reprocessed using one of the other algorithms implemented in the pipeline. These include the following: (a) Mean value of the background pixels; (b) Mode of the background pixels; (c) The original background estimation procedure described in \cite{King2014}, which searches for the centroid in the sigma-clipped and smoothed histogram of background pixels values. 

\subsubsection{Aperture photometry} \label{subsubsec:twoap}

The full width at half maximum (FWHM) of the vertical pair of spots is different from that of the horizontal pair (see Fig.~\ref{fig:rbpl_src}), due to the instrument design. The optimal size of the photometry aperture, which minimizes the uncertainties, depends on the FWHM. Therefore, the current version of the pipeline measures and treats two pairs of spots independently.

For AGN with prominent host galaxies, measured polarization values are reduced by unpolarized emission of the host inside the aperture. This depolarization effect varies with seeing: fractional polarization is lower for shots with larger FWHM \citep{Sosa2017}. In order to minimize this effect for sources with bright hosts we used fixed apertures 3~arcsec to 6~arcsec in diameter depending on the source. The corresponding value for each source is given in the 'aperture' column of Table~\ref{tab:sample}. All other unresolved objects or sources with negligible host galaxy contribution were measured with aperture dependent on FWHM that optimised SNR. We processed each source with 8 different apertures with sizes in the range 1 -- 4.5 $\times$ FWHM. Then we visually verified that the fractional polarization and EVPA stabilised within this range of apertures. Finally, we selected the aperture size that provided the highest SNR in fractional polarization in the stability range. This value was 5.1~arcsec on median.

\subsubsection{Centroids of the main target spots} \label{subsubsec:centfail}
\begin{figure}
 \centering
 \includegraphics[width=0.2\textwidth]{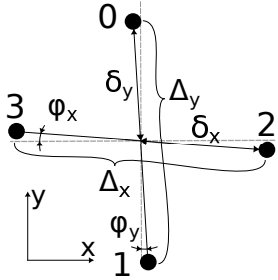}
 \caption{The pattern of 4 spots at each position $(x,y)$ on the CCD and 6 parameters describing it. $\Delta_x(x,y)$ is the distance between the horizontal spots, $\delta_x(x,y)$ is the distance from the right-spot to the central point, and $\phi_x(x,y)$ is the angle between the CCD x-axis and the line connecting horizontal points. $\Delta_y(x,y)$, $\delta_y(x,y)$ and $\phi_y(x,y)$ denote similar quantities for the vertical pair of spots.}
 \label{fig:spot_diagr}
\end{figure}

The pipeline uses SExtractor \citep{Bertin1996} for identification of sources in the image. It provides a windowed centroid position for each source in the frame. In a small fraction of images with a strong gradient of the sky background at the edges of the masked regions, the centroid calculation procedure fails. This causes inaccurate 
polarization measurements of the central target. In order to avoid this problem, we have added a second procedure for centroid computation, based on the PyGuide\footnote{\url{http://staff.washington.edu/rowen/PyGuide/Manual.html}} library. After the identification of the central target spots in the SExtractor output catalogue, the pipeline finds parameters $\Delta_x$, $\delta_x$, $\phi_x$, $\Delta_y$, $\delta_y$ and $\phi_y$ of the spatial distortions model, whose meaning is demonstrated in Fig.~\ref{fig:spot_diagr} (see also \S 2.2.1 in \cite{King2014}). If any of these parameters deviates from the model prediction more than the maximum residual in the model fit (see \S~\ref{subsec:model} and Fig.~\ref{fig:spatial_model:delx}, \ref{fig:spatial_model:dx}), then the spots centroids calculation is considered to have failed. In this case the second PyGuide based centroid computation algorithm is used. The combination of both algorithms has proven sufficient in practice. 
We have not identified any image in our observations where both methods fail.

\subsubsection{Additional astrometry procedure} \label{subsubsec:astr2}

The {\em RoboPol} pipeline uses the Astrometry.net \citep{Lang2010} library to perform astrometry. Accurate coordinates of the field sources are needed for the pointing procedure during observations and for the differential photometry during the data processing. In the case of a bright sky or very sparse stellar field \citep[e.g. in some of the fields studied in][or AGN at high galactic latitudes]{Skalidis2018} Astrometry.net is unable to find a solution for the World Coordinate System. For this reason we have added a second astrometry routine which is based on the Alipy\footnote{\url{https://obswww.unige.ch/~tewes/alipy/}} library. This routine
starts only if Astrometry.net fails. It finds a geometrical transformation between the current frame and a preliminary stored source catalogue, which has a World Coordinate System (WCS) solution. Then  the calculated transformation in the CCD frame is converted to WCS transformation and, thereby, the new WCS of the current frame is defined. This routine requires either a previous {\em RoboPol} image of the same field solved by Astrometry.net or the Digitized Sky Survey fits image with a WCS solution in the header.

\subsubsection{Polarization measurements statistics} \label{subsubsec:evpaunc}

Values of PD and its uncertainty were calculated following Eq.~5 in \cite{King2014} under the assumption of Gaussianity of the Stokes parameters. Any linear polarization measurement is subject to bias towards higher polarization degree (PD) values \citep{Serkowski1958}. The PD follows the \cite{Rice1945} distribution, which  deviates significantly from the normal distribution at low SNR. Since there is a variety of methods suggested for correction of this bias \citep[e.g.][]{Simmons1985,Vaillancourt2006,Plaszczynski2014} we did not include any bias correction of the fractional polarization into the pipeline. {\bf The monitoring data discussed in this paper is uncorrected for bias.}

EVPA measurements are also non-Gaussian and defined by the following probability density 
\citep{Naghizadeh1993}:
\begin{equation}
\begin{multlined}
 G({\rm EVPA,EVPA_0,PD_0}) = \\
 = \frac{1}{\sqrt{\pi}} \left\{ \frac{1}{\sqrt{\pi}} + \eta_0 \exp\left(\eta_0^2\right)
 [ 1 + \erf(\eta_0) ] \right\} \exp\left(-\frac{\rm PD_0^2}{2 \sigma_{\rm PD}^2}\right),
\end{multlined}
\end{equation}
where $\eta_0 = {\rm PD_0}\cos{2({\rm EVPA} - {\rm EVPA_0})}/(\sigma_{\rm PD} \sqrt{2})$, $\erf$ is the 
Gaussian error function, ${\rm PD_0}$ and ${\rm EVPA_0}$ are the true values of PD and EVPA and $\sigma_{\rm PD}$ is 
the uncertainty of PD.

In the latest version of the pipeline we determine the EVPA uncertainty $\sigma_{\rm EVPA}$ 
numerically solving the following integral:
\begin{equation}
 \int_{-1\sigma_{\rm EVPA}}^{1\sigma_{\rm EVPA}} G({\rm EVPA,PD_0}) {\rm dEVPA} = 0.6827.
\end{equation}
The true PD value in this procedure is estimated using the Modified ASymptotic (MAS) estimator proposed by \cite{Plaszczynski2014} as follows:
\begin{equation}
{\rm PD_0} = {\rm PD} - \sigma_{\rm PD}^2 \frac{1 - \exp(-{\rm PD}^2/\sigma_{\rm PD}^2)}{\rm 2 PD}
\end{equation}
For high SNR values ${\rm PD}/\sigma_{\rm PD} \ge 10$ the uncertainty of EVPA is approximated as $\sigma_{\rm EVPA} = 0.5 \sigma_{\rm PD} / {\rm PD_0}$.

\begin{figure}
 \centering
 \includegraphics[width=\columnwidth]{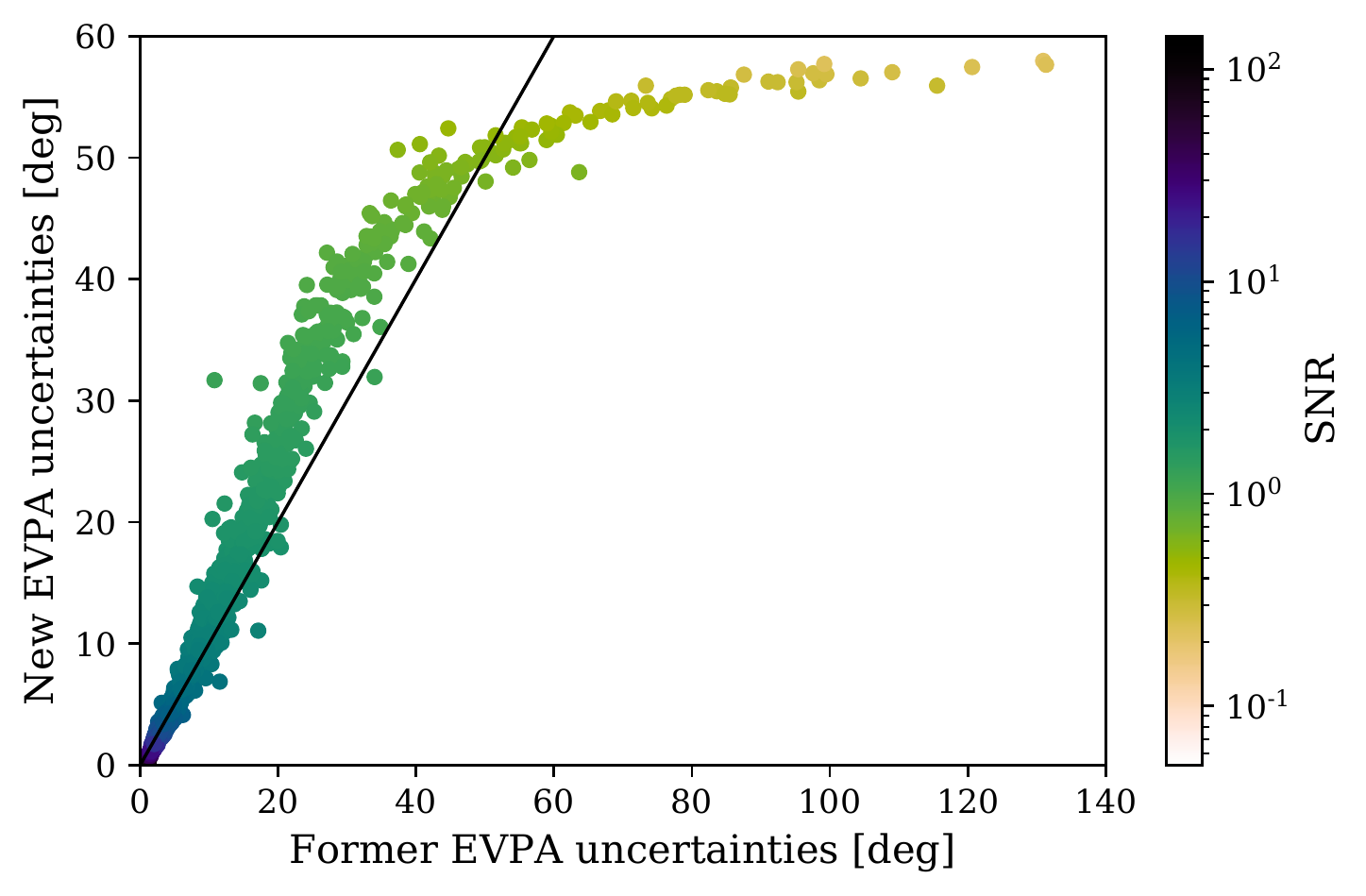}
 \caption{New EVPA uncertainty estimates calculated according to \S~\ref{subsubsec:evpaunc} plotted against the former estimates calculated according to \protect\cite{King2014}. The solid line is y=x. The colour coding shows the signal-to-noise-ratio of the polarization fraction.}
 \label{fig:evpaunc}
\end{figure}
Figure~\ref{fig:evpaunc} shows the new EVPA uncertainty estimates plotted against the former estimates calculated with the previous pipeline as described in \cite{King2014}. The uncertainties tend to have been underestimated in moderately low SNR ($1 \lesssim {\rm SNR} \lesssim 10$) and to have been overestimated in very low ${\rm SNR} < 1$.

\subsection{Instrument model} \label{subsec:model}

Instrumental polarization, vignetting and spatial distortions of the four-spots-pattern are corrected by the instrument model. The instrument model consists of two independent parts. The first part approximates 
the dependencies of the six parameters in Fig.~\ref{fig:spot_diagr} on the (x,y)-position on the CCD. The second part is a spatial function that describes multipliers for photon counts in each of the four spots as a function of (x,y). These multipliers compensate for the global variations of the instrumental polarization in the field and vignetting. The instrument model coefficients are obtained by fitting  the model to measurements from a raster scan of a zero-polarized star in the field. The complete description of the {\em RoboPol} instrument model was presented by \cite{King2014}.

The instrument response can vary with time, telescope position, flexure, temperature etc. For this reason, we performed a raster scan of a zero-polarized star several times per season and after every removal of the {\em RoboPol} polarimeter from the telescope. During the 2013 -- 2017 period we obtained 11 model scans using 7 different zero-polarized standards in different positions of the telescope. Comparing results of the model fit to these data, we do not find any significant systematic difference between these models. 

Combining multiple model raster scans of zero-polarized stars that were observed at different epochs, we are able to reduce the random errors of the model fits. Therefore, for the data discussed in this paper we used the combined model. It was created using 1624 measurements of unpolarized stars in all 11 raster scans obtained in 2013 -- 2017 and shown in Fig.~\ref{fig:raster_map_plot} of Appendix~\ref{ap:b}. Figures~\ref{fig:spatial_model:delx} -- 
\ref{fig:qu_res_hist} of the same appendix show the fits of model parameters and their residuals. This model gives the best approximation of the instrumental polarization and is less noise-dominated than any single raster scan.

\section{Calibration using standard stars} \label{sec:polar}

The instrument model accounts for the variation of the polarization response of the instrument globally across the FoV, while all AGN were measured in the narrow central masked area. Since the model can deviate locally from the real instrumental polarization value we have also been monitoring polarimetric standard stars in the mask. The list of stars observed during the project execution period and their catalogued polarization values together with corresponding references are given in Table~\ref{tab:pol_stnds}.
\begin{table*}
\centering
\caption{Polarization parameters of standard stars monitored by {\em RoboPol}, as reported in the literature.}
\label{tab:pol_stnds}
\begin{tabular}{lcccc}
\hline
Source & Band & PD (\%) &  EVPA (\degr) & Reference \\
\hline
\multicolumn{5}{c}{polarized} \\
BD$+$57.2615    & R  & 2.02   $\pm$ 0.05  & 41.0 $\pm$ 1.0  & \cite{Whittet1992}\\
BD$+$59.389     & R  & 6.430 $\pm$ 0.022 & 98.14 $\pm$ 0.10 & \cite{Schmidt1992}\\
BD$+$64.106     & R  & 5.150 $\pm$ 0.098 & 96.74 $\pm$ 0.54 & \cite{Schmidt1992}\\
CMaR1 24        & R  & 3.18  $\pm$ 0.09  & 86.0 $\pm$ 1.0  & \cite{Whittet1992}\\
CygOB2 14       & R  & 3.13  $\pm$ 0.05  & 86.0 $\pm$ 1.0  & \cite{Whittet1992}\\
\multirow{2}{*}{HD147283 $\left\lbrace\begin{array}{l}{}\\{}\end{array}\right.$} & R  & 1.59  $\pm$ 0.03  & 174.0 $\pm$ 1.0  & \cite{Whittet1992}\\
          & R  &   1.81      &     176.0   & \cite{Carrasco1973}\\
HD147343        & R  & 0.43  $\pm$ 0.05  & 151.0 $\pm$ 3.0  & \cite{Whittet1992}\\
HD150193        & R  & 5.19  $\pm$ 0.05  & 56.0  $\pm$ 1.0  & \cite{Whittet1992}\\
\multirow{2}{*}{HD154445 $\left\lbrace\begin{array}{l}{}\\{}\end{array}\right.$} & R  & 3.683 $\pm$ 0.072 & 88.91 $\pm$ 0.56 & \cite{Schmidt1992}\\
\               & R  & 3.63  $\pm$ 0.01  & 90.0  $\pm$ 0.1  & \cite{Hsu1982}\\
HD155197        & R  & 4.274 $\pm$ 0.027 & 102.88 $\pm$ 0.18 & \cite{Schmidt1992}\\
HD161056        & R  & 4.012 $\pm$ 0.032 & 67.33 $\pm$ 0.23 & \cite{Schmidt1992}\\
\multirow{2}{*}{HD183143${}^b$ $\left\lbrace\begin{array}{l}{}\\{}\end{array}\right.$} & R  & 5.90  $\pm$ 0.05  & 179.2 $\pm$ 0.2  & \cite{Hsu1982}\\
                & R  & 5.7   $\pm$ 0.04  & 178.0 $\pm$ 1.0  & \cite{Bailey1982}\\
\multirow{3}{*}{HD204827${}^b$ $\left\lbrace\begin{array}{l}{}\\{}\\{}\end{array}\right.$} & R  & 4.893 $\pm$ 0.029 & 59.10 $\pm$ 0.17 & \cite{Schmidt1992}\\
                & R  & 4.86  $\pm$ 0.05  & 60.0  $\pm$ 1.0  & \cite{Bailey1982}\\
                & R  & 4.99  $\pm$ 0.05  & 59.9  $\pm$ 0.1  & \cite{Hsu1982}\\
HD215806        & R  & 1.83  $\pm$ 0.04  & 66.0  $\pm$ 1.0  & \cite{Whittet1992}\\
HD236633        & R  & 5.376 $\pm$ 0.028 & 93.04 $\pm$ 0.15 & \cite{Schmidt1992}\\
Hiltner960${}^a$& R  & 5.210 $\pm$ 0.029 & 54.54 $\pm$ 0.16 & \cite{Schmidt1992}\\
\multirow{3}{*}{VICyg12${}^b$ $\left\lbrace\begin{array}{l}{}\\{}\\{}\end{array}\right.$}   & R  & 7.97  $\pm$ 0.05  & 117.0 $\pm$ 1.0  & \cite{Whittet1992}\\
                & R  & 7.893 $\pm$ 0.037 & 116.23 $\pm$ 0.14 & \cite{Schmidt1992}\\
                & R  & 7.18  $\pm$ 0.04  & 117.0 $\pm$ 1.0  & \cite{Hsu1982}\\
\multicolumn{5}{c}{unpolarized} \\
BD$+$28.4211    & V  & 0.054 $\pm$ 0.027 &        54.22       & \cite{Schmidt1992}\\
BD$+$32.3739    & V  & 0.025 $\pm$ 0.017 &        35.79       & \cite{Schmidt1992}\\
BD$+$33.2642    & R  & 0.20  $\pm$ 0.15  &   78    $\pm$ 20   & \cite{Skalidis2018}\\
BD$+$40.2704    & ?  & 0.07  $\pm$ 0.02  &   57    $\pm$ 9    & \cite{Berdyugin2002}\\
G191B2B         & V  & 0.061 $\pm$ 0.038 &       147.65       & \cite{Schmidt1992}\\
HD14069         & V  & 0.022 $\pm$ 0.019 &       156.57       & \cite{Schmidt1992}\\
HD154892        & B  & 0.05  $\pm$ 0.03  &         -          & \cite{Turnshek1990}\\
HD212311        & V  & 0.034 $\pm$ 0.021 &       50.99        & \cite{Schmidt1992}\\
HD21447         & V  & 0.051 $\pm$ 0.020 &      171.49        & \cite{Schmidt1992}\\
HD94851         & B  & 0.057 $\pm$ 0.018 &         -          & \cite{Turnshek1990}\\
WD2149+021      & R  & 0.050 $\pm$ 0.006 & $-$63   $\pm$ 3    & \cite{Cikota2017}\\
 \hline
 \multicolumn{5}{l}{${}^a$ - possibly variable \cite{You2017};${}^b$ - variable in {\em RoboPol} data or/and in \cite{Hsu1982,Dolan1986};}\\
 \end{tabular}
\end{table*}
We note that some stars monitored in the project turned out to be variable, despite the fact that they are considered as standards in other polarimetric programmes and literature. For example, we certainly observe variability in HD~204827, HD~183143 and VI~Cyg~12, which is in agreement with \cite{Hsu1982,Dolan1986}. Moreover, some standards appear to be stable but their polarization significantly deviates from that reported in the literature (e.g. CMaR1 24 and BD$+$57.2615). Furthermore, the situation is complicated by multiple inconsistent polarimetric parameters reported in different works for some standards. In general, the situation with optical polarimetric standards can be characterised as disheartening. For this reason in 2018 we started an effort intended to establish a well defined sample of stable polarimetric standards. Its results will be reported elsewhere.

\begin{figure}
 \centering
 \includegraphics[width=0.44\textwidth]{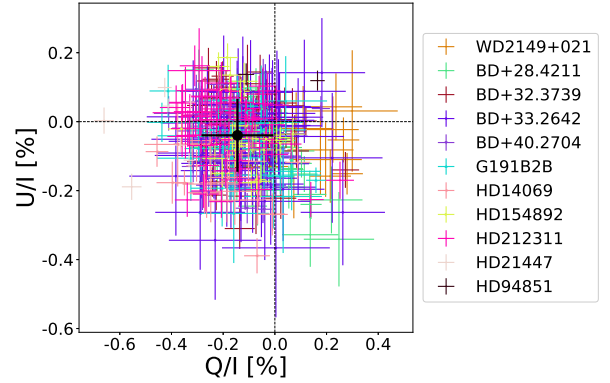}
 \caption{Relative Stokes parameters of zero-polarization standards observed in 2013 -- 2017. The weighted mean is shown by the black point.}
 \label{fig:qu_unp}
\end{figure}

Figure \ref{fig:qu_unp} shows the distribution of relative Stokes parameters of zero-polarized standards measured along the whole observing period. Measurements of WD2149+021 systematically deviate from the centroid of all other measurements. Therefore, this star was excluded from the analysis. The weighted mean centroid for all other measurements in Fig.~\ref{fig:qu_unp} is located at ${\rm Q/I}=-0.0015$, ${\rm U/I}=-0.0004$. This value is the difference between the model-predicted and the real instrumental polarization at the central mask region.
We used this quantity as an additional instrument polarization correction for all AGN measurements. The standard deviations of the Stokes parameters in Fig.~\ref{fig:qu_unp} are $0.0014$ and $0.0011$. These values were considered as uncertainties and propagated to the AGN measurements uncertainties. It is worth noting that in this work we use a different approach from the one reported in \cite{instrumentpaper}, where polarization standards were processed without the instrument model correction. There, it was found that the instrumental polarization varies between observing seasons with an amplitude $\sim$ 0.1 -- 0.2 per cent. This variation can be partly explained by the fact that the mask shadow location on the CCD changes slightly with time. Unlike \cite{instrumentpaper}, in this work we use the instrument model correction of polarization parameters, which accounts for changing position of a target on the CCD. Presumably for this reason we do not find any significant variability of the instrumental polarization among observing seasons. Thus, here we use the constant instrumental polarization correction value for all measurements. Even if a residual systematic error is present in such an instrumental polarization correction, its value is negligible compared to typical photon noise of our AGN measurements. 

\begin{figure}
 \centering
 \includegraphics[width=0.3\textwidth]{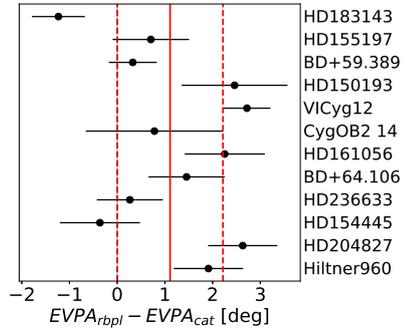}
 \caption{Differences between weighted average of observed EVPA and corresponding catalogue values for 12 most reliable highly-polarized standards. Weighted mean value for all 12 stars is shown by the solid red line, while their weighted standard deviation is shown by the red dashed lines.}
 \label{fig:pa_diff}
\end{figure}

In order to find the rotation of the instrumental Q/I - U/I plane with respect to the standard reference frame we used 331 observations of 12 high-polarization stars that are considered to be the most stable standards. The difference between catalogued and averaged {\em RoboPol} values of EVPA for these stars are shown in Fig.~\ref{fig:pa_diff}. The weighted average  for 12 stars is ${\rm EVPA_{rbpl}} - {\rm EVPA_{cat}} = 1.1\pm1.1\dg$. Relative Stokes parameters Q/I and U/I of AGN were corrected for this EVPA zero point offset and the uncertainties were propagated.

\section{Notes on individual sources} \label{sec:ind_src}

In this section we list caveats that arose during the processing of individual sources in the sample.\newline
\textbf{J0035+5950:} A source of comparable brightness is located 1.58~arcsec from the blazar. It was initially considered as the lensed image of the blazar, however it was not detected in radio \citep{2015Aleksic}. Due to the small separation between the two sources, it could not be resolved in all seeing conditions in {\em RoboPol} images. For this reason we performed the aperture photometry with a fixed 6~arcsec aperture surrounding both sources. Therefore, the polarization fraction values are most likely significantly underestimated, while the EVPA should be reliable.\newline
\textbf{J0324+3410:} For this source we used a fixed 4~arcsec aperture. However, this Narrow-Line Seyfert 1 (NLSy1) has asymmetric spiral arms resolved in the host galaxy \citep{Zhou2007}. The spiral arms can potentially give non-zero polarization of the stellar light. A dedicated study is needed to clarify whether the starlight of the host has significant polarization.\newline
\textbf{J0728+5701:} A foreground star is located nearby. To avoid depolarization due to contamination by the star light we used a fixed 4~arcsec aperture.\newline
\textbf{J0849+5108:} This NLSy1 has a prominent spiral host galaxy \citep{Hamilton2020}. It appears to be asymmetric in images and can potentially contribute to the polarization. We measured the nucleus with a fixed 4~arcsec aperture.\newline
\textbf{J1148+5924:} NGC 3894 has a very bright host. The galaxy fills the masked areas entirely. For this reason there is a possibility of deviation of the PD from its real values due to inaccurate estimate of the sky background. We measured this source with a fixed 6~arcsec aperture.\newline
\textbf{J1442+1200:} 1ES 1440+122 has a prominent host galaxy. There is also a nearby source located 2.5~arcsec from the core according to Gaia data \citep{GaiaDR2}. It could potentially affect the polarization. We measured the nucleus with a fixed 4.5~arcsec aperture.\newline
\textbf{J1505+0326:} This NLSy1 has a prominent elliptical host galaxy, however there is an asymmetric structure 3~arcsec from the core that is probably produced by an interaction with another galaxy \citep{2018DAmmando}. Since we measured the source with a fixed 4~arcsec aperture the emission of this component can potentially contribute to the polarization.\newline
\textbf{J1653+3945:} Mkn501 has a very bright host. The galaxy fills the masked areas entirely. For this reason there is a possibility of deviation of the PD from its real values due to inaccurate estimate of the sky background. We measured this source with a fixed 6~arcsec aperture.\newline
\textbf{J1728+0427:} PKS 1725+044 has a prominent host galaxy. There is also a nearby source located 1.6~arcsec from the core according to Gaia data. It could potentially affect the polarization. We measured the nucleus with a fixed 4~arcsec aperture.\newline
\textbf{J1733$-$1304:} PKS 1730$-$130 has a nearby source located 1.8~arcsec from the core according to Gaia data. It could potentially affect the polarization. We measured the nucleus with a fixed 5~arcsec aperture.\newline
\textbf{J1743+1935:} This AGN has a prominent host galaxy. There is also a nearby source located 3.4~arcsec from the core according to Gaia data. It could potentially affect the polarization. We measured the nucleus with a fixed 5~arcsec aperture.\newline
\textbf{J1943+2118:} A few measurements for this TeV source have been presented in \citep{Hovatta2016}. However, reanalysing data for this paper, we discovered that the source was previously misassociated. The measured polarization corresponds to a $\sim 17$ mag star, while the TeV AGN is associated with 2MASS source J1943562+2118233 \citep{Landi2009} that is located 5.3~arcsec North-West from this star. Since the AGN is very faint $R\approx22.4$~mag, it cannot be measured in our images. For this reason J1943+2118 is not present in the data table.\newline
\textbf{J2031+1219:} There is a nearby source located 2.9~arcsec from the core according to Gaia data. It could potentially affect the polarization. We measured the nucleus with a fixed 3~arcsec aperture.\newline
\textbf{J2033+2146:} This AGN has a prominent host galaxy. There is also a nearby foreground star located 2.3~arcsec from the core according to Gaia data. It could potentially affect the polarization. We measured the nucleus with a fixed 4~arcsec aperture.\newline

\section{Monitoring data and polarization parameters} \label{sec:data}

The data products that are made publicly available in this work are the
polarimetric monitoring data of the sample sources and their average polarization parameters. Machine-readable form of the data tables can be accessed via Harvard dataverse\footnote{\url{https://doi.org/10.7910/DVN/IMQKSE}} and Vizier. Abridged versions of the average polarization parameters table and monitoring data table are also presented in the Appendix~\ref{ap:a}.

In Table~\ref{tab:sample} of Appendix~\ref{ap:a} we list general information about AGN in the sample including their equatorial coordinates, redshift with corresponding reference and their sample ID.
We also give the total number of observations of each source, and the number of seasons during which it has been observed, and the median time sampling interval.
Additionally, we provide the following statistics to quantify the average and variability of the fractional polarization and the EVPA. We estimate the intrinsic mean polarization fraction $p_0$ and the intrinsic modulation index $m_p$ by modelling the distribution of the measured polarization fraction as a Beta distribution following \cite{Blinov2016a}. We infer the distribution parameters with Bayesian modelling using \verb|PyStan|\footnote{\url{https://pystan.readthedocs.io/}}. To quantify the EVPA variability we take the difference of the 75 per cent and 25 per cent quantiles of the measured EVPA distribution. We subtract this value from $180^\circ$ if the difference exceeds $90^\circ$. The result is divided by $90^\circ$. This variability index, $v_\chi$, is normalised to the interval $[0,1]$, where 0 corresponds to a perfectly stable EVPA and 1 corresponds to a variable EVPA with a perfectly uniform distribution, i.e. no preferred orientation. For sources with $v_\chi < 0.5$ we show the wrap-corrected median EVPA as a measure of the preferred orientation.

The entire AGN monitoring data are given in electronic format, while the first rows of this table are given in Table~\ref{tab:mon_dat}. There for each observation we provide Julian Date (JD), relative Stokes parameters and their uncertainties before the instrument polarization correction, relative Stokes parameters after the instrument polarization and EVPA zero-point correction as well as corresponding values of PD and EVPA.

\section{Data Policy}\label{sec:datapolicy}
These data are being made available to the public as a service to the astronomical community. If you use {\em RoboPol} data in your research, we request that you cite the present publication, allowing us to keep track of the impact of our work, and that you include the following acknowledgement: 'This research has made use of data from the {\em RoboPol} programme, a collaboration between Caltech, the University of Crete, IA-FORTH, IUCAA, the MPIfR, and the Nicolaus Copernicus University, which was conducted at Skinakas Observatory in Crete, Greece.'

\section*{Acknowledgements}
We thank A. Steiakaki and E. Paleologou for their invaluable contribution
and technical support. The {\em RoboPol} project is a collaboration between Caltech in the USA, MPIfR in Germany, Toru\'{n} Centre for Astronomy in Poland, the University of Crete/FORTH in Greece, and IUCAA in India. The U. of Crete group acknowledges support by the ``RoboPol'' project, which was implemented under the ``Aristeia'' Action of the  ``Operational Programme Education and Lifelong Learning'' and was co-funded by the European Social Fund (ESF) and Greek National Resources, and by the European Comission Seventh Framework Programme (FP7) through grants PCIG10-GA-2011-304001 ``JetPop'' and PIRSES-GA-2012-31578 ``EuroCal''. This research was supported in part by NASA grant NNX11A043G and NSF grant AST-1109911, and by the Polish National Science Centre, grant  numbers 2011/01/B/ST9/04618 and 2017/25/B/ST9/02805. C.C., D.B., N.M., R.S. and K.T. acknowledge support from the European Research Council (ERC) under the European Union Horizon 2020 research and innovation programme under the grant agreement No 771282. K.T. and G.V.P. acknowledge support by the European Commission 
Seventh Framework Programme (FP7) through the Marie Curie Career Integration Grant PCIG-GA-2011-293531 ``SFOnset'' as well as NASA Hubble Fellowship grant  \#  HST-HF2-51444.001-A awarded  by  the  Space Telescope  Science  Institute, which  is  operated  by  the Association of Universities for Research in Astronomy, Incorporated, under NASA contract NAS5-26555. T.H. was supported by the Academy of Finland projects 317383, 320085, and 322535

\section*{DATA AVAILABILITY}

The data underlying this article are available in Harvard Dataverse, at \url{https://dx.doi.org/10.7910/DVN/IMQKSE}.

%%%%%%%%%%%%%%%%%%%%%%%%%%%%%%%%%%%%%%%%%%%%%%%%%%



%\bibliographystyle{mnras}

\bibliographystyle{mnras}
% Use the LaTeX power, use bibtex properly.
\bibliography{references,tableA1.bib}

%%%%%%%%%%%%%%%%%%%%%%%%%%%%%%%%%%%%%%%%%%%%%%%%%%
\appendix

\section{Sample sources information and monitoring data}
\label{ap:a}

Table~\ref{tab:sample} lists the monitored sources, additional source information, and polarization statistics as described in \S~\ref{sec:data}. 
%Fig.~\ref{fig:sampling_stats} shows the distributions of the sampling statistics for all sources. The distributions of the intrinsic mean and modulation index of the fractional polarization are shown in Fig.~\ref{fig:pd_stats}. Fig.~\ref{fig:evpa_stats} shows the distribution of the EVPA variability.

\clearpage
\onecolumn
\begin{landscape}
{\scriptsize
\begin{table*}
\centering
\caption{Information on the sample sources. (1) - J2000 name; (2) - Alternative source identifier; (3) - Right ascension; (4) - Declination; (5) - Redshift; (6) - Redshift reference; (7) - Sample identifier; (8) - Aperture; (9) - Number of measurements; (10) - Number of seasons; (11) - Median time sampling interval; (12) - Mean intrinsic polarisation fraction; (13) - Intrinsic modulation index of the polarization fraction; (14) - Variability index of the EVPA; (15) - Preferred EVPA orientation of less variable sources. The entire table is available online.}
\label{tab:sample}
\begin{tabular}{llrrrrrrrrrrrrr}
\hline
Name     & Alt. name  & RA (J2000)  & DEC (J2000)  & $z$ $^a$   & $z$ ref.  & SID$^b$   & Aperture$^c$  & $N$ meas.  & $N$ seas.  & sampling  & $p_0$  & $m_p$  & $v_\chi$  & $\chi_0$  \\
   &            & h:m:s       & \degr:\arcmin:\arcsec        &       &           &           &  arcsec       &          & d         & d     &  \%      &    &       & deg  \\
(1)      & (2)        & (3)         & (4)          &  (5)  & (6)       & (7)       & (8)           & (9)      & (10)     & (11)      & (12)        & (13)   & (14)   & (15) \\
\hline
 RBPLJ0006$-$0623 &      PKS 0003$-$066 &  00:06:13.9 &  $-$06:23:35.3 &                     0.34668 &  \cite{2009MNRAS.399..683J} &      5 &    v &   15 &  2 &   8 &  21.20 &  0.45 &  0.13 & $-$14.0 \\
 RBPLJ0017+8135 &        S5 0014+81 &  00:17:08.5 &  $+$81:35:08.1 &                       3.366 &  \cite{1994ApJ...436..678O} &    2,4 &    v &   30 &  3 &   7 &   0.80 &  0.62 &  0.65 &     - \\
 RBPLJ0035+5950 &      1ES 0033+595 &  00:35:52.7 &  $+$59:50:04.2 &                       0.086 &  \cite{1999AnA...352...85F} &      5 &    6 &    5 &  1 &   8 &      - &     - &     - &     - \\
 RBPLJ0045+2127 &                 - &  00:45:19.3 &  $+$21:27:40.0 &                           - &                           - &      3 &    v &   64 &  4 &   3 &   4.83 &  0.59 &  0.84 &     - \\
 RBPLJ0102+5824 &      TXS 0059+581 &  01:02:45.8 &  $+$58:24:11.1 &                       0.644 &  \cite{2005ApJ...626...95S} &      5 &    v &   18 &  3 &   9 &  13.45 &  0.32 &  0.53 &     - \\
 RBPLJ0114+1325 &                 - &  01:14:52.8 &  $+$13:25:37.5 &    0.583\textsuperscript{*} &  \cite{2014ApJ...784..151S} &      3 &    v &   59 &  4 &   4 &   7.04 &  0.48 &  0.74 &     - \\
 RBPLJ0136+3905 &       B3 0133+388 &  01:36:32.5 &  $+$39:05:59.6 &                       0.750 &  \cite{2015AnA...575A..21N} &      6 &    v &    7 &  2 &  27 &      - &     - &     - &     - \\
 RBPLJ0136+4751 &            OC 457 &  01:36:58.6 &  $+$47:51:29.0 &                       0.859 &  \cite{1987ApJS...63....1H} &      3 &    v &   64 &  4 &   3 &   8.75 &  0.64 &  0.86 &     - \\
 RBPLJ0152+0146 &          RBS 0248 &  01:52:39.6 &  $+$01:47:17.4 &                        0.08 &  \cite{1998ApJS..118..127L} &      6 &    4 &    9 &  2 &  20 &      - &     - &     - &     - \\
 RBPLJ0211+1051 &                 - &  02:11:13.2 &  $+$10:51:35.0 &      0.2\textsuperscript{*} &  \cite{2010ApJ...712...14M} &      3 &    v &   65 &  4 &   3 &  14.98 &  0.51 &  0.31 & $-$30.0 \\
 RBPLJ0217+0837 &                 - &  02:17:17.1 &  $+$08:37:03.9 &                       0.085 &  \cite{2013ApJ...764..135S} &      3 &    v &   62 &  4 &   3 &   6.11 &  0.47 &  0.50 &  76.0 \\
 RBPLJ0221+3556 &        S4 0218+35 &  02:21:05.5 &  $+$35:56:13.9 &                       0.944 &  \cite{2003ApJ...583...67C} &      5 &    v &    1 &  1 &   - &      - &     - &     - &     - \\
 RBPLJ0222+4302 &            3C 66A &  02:22:39.6 &  $+$43:02:07.8 &                        0.37 &  \cite{2013ApJ...766...35F} &    5,6 &    v &   25 &  2 &   6 &   7.99 &  0.40 &  0.38 &  16.0 \\
 RBPLJ0232+2017 &                 - &  02:32:48.6 &  $+$20:17:17.4 &                        0.14 &  \cite{1993ApJ...412..541S} &      6 &    4 &    8 &  2 &  18 &      - &     - &     - &     - \\
 RBPLJ0238+1636 &       AO 0235+164 &  02:38:38.9 &  $+$16:36:59.0 &                        0.94 &  \cite{1987ApJS...63....1H} &      5 &    v &   20 &  3 &   5 &  12.15 &  0.49 &  0.39 &  22.0 \\
 RBPLJ0259+0747 &      PKS 0256+075 &  02:59:27.1 &  $+$07:47:39.6 &                       0.893 &  \cite{1993MNRAS.264..298M} &      3 &    v &   34 &  3 &   5 &  23.59 &  0.47 &  0.07 &  53.0 \\
 RBPLJ0303+4716 &       B3 0300+470 &  03:03:35.2 &  $+$47:16:16.3 &                       0.475 &  \cite{1992ApJ...396..469H} &      5 &    v &   15 &  3 &  10 &   7.25 &  0.45 &  0.40 &  81.0 \\
 RBPLJ0303$-$2407 &      PKS 0301$-$243 &  03:03:26.5 &  $-$24:07:11.5 &                      0.2657 &  \cite{2014AnA...565A..12P} &      3 &    v &   12 &  3 &  17 &   6.67 &  0.24 &  0.19 &  54.0 \\
 RBPLJ0316+4119 &      TXS 0313+411 &  03:16:43.0 &  $+$41:19:29.9 &                     0.01894 &  \cite{2002AJ....123.2990B} &      5 &    4 &    8 &  2 &  19 &      - &     - &     - &     - \\
 RBPLJ0319+1845 &          RBS 0413 &  03:19:51.8 &  $+$18:45:34.4 &                        0.19 &  \cite{1991ApJS...76..813S} &    5,6 &    v &    7 &  3 &  12 &      - &     - &     - &     - \\
\multicolumn{15}{l}{\dots} \\
\hline
\multicolumn{15}{p{0.95\textwidth}}{${}^a$ - Redshift notes: (*) non spectroscopic redshift, (l) lower limit on redshift, (m) multiple values are present in literature. ${}^b$ - Sample IDs: (1) \g-ray--loud sub-sample from the 'June survey' sample, (2) \g-ray--quiet sub-sample from the 'June survey' sample, (3) \g-ray--loud sub-sample from the monitoring sample, (4) \g-ray--quiet sub-sample from the monitoring sample, (5) additional sample of 68 hand-picked sources of high interest, (6) ISP/HSP blazar sample. ${}^c$ - fixed aperture is given in arcsec, 'v' denotes variable aperture (for details see sections \ref{subsubsec:twoap} and \ref{sec:ind_src}).} \\
\end{tabular}
\end{table*}
}
\end{landscape}
\twocolumn

\begin{table*}
\centering
\caption{Monitoring data: (1) - J2000 name; (2) - Julian Date; (3,4) - Q/I, U/I relative Stokes parameters before the correction for the instrumental polarization; (5,6) - Q/I and U/I relative Stokes parameters corrected for the instrumental polarization and EVPA zero-point; (7,8) - Fractional polarization and polarization vector position angle corrected for the instrumental polarization and EVPA zero-point. This table gives only first 3 rows of the entire dataset, which can be accessed at \url{https://doi.org/10.7910/DVN/IMQKSE}.}
\label{tab:mon_dat}
  \begin{tabular}{lcccccS[table-format=2.1(1)]S[table-format=2.1(1)]} 
  \hline
 AGN ID &   JD   & ${\rm Q/I}_{\rm inst}$ & ${\rm U/I}_{\rm inst}$ &    Q/I   & U/I  &  \si{PD}  &  \si{EVPA} \\
        &        &                        &                        &          &      & \si{\%} &  \si{deg}  \\
   (1)  &   (2)  &     (3)                &      (4)               &    (5)   & (6)  & \si{(7)}  &  \si{(8)}  \\
 \hline
RBPLJ0017+8135    &  2451023.41611  &  $0.0131\pm0.00022$  &  $0.0121\pm0.00012$  &  $0.0131\pm0.00022$  &  $0.0121\pm0.00012$  & 5.2(3) & 20.4(12) \\
RBPLJ0017+8135    &  2459016.01214  &  $0.0031\pm0.00002$  &  $0.0048\pm0.00005$  &  $0.0431\pm0.00022$  &  $0.1221\pm0.00022$  & 12.2(1) & 20.4(12) \\
RBPLJ0017+8135    &  2459086.08916  &  $0.0536\pm0.00003$  &  $0.0022\pm0.00006$  &  $0.0321\pm0.00022$  &  $0.0441\pm0.00043$  & 6.2(2) & 120.4(21) \\
\multicolumn{8}{l}{\dots} \\
\hline
\end{tabular}
\end{table*}

\clearpage

\section{Instrument model} \label{ap:b}

\begin{figure}
 \centering
 \includegraphics[width=7cm]{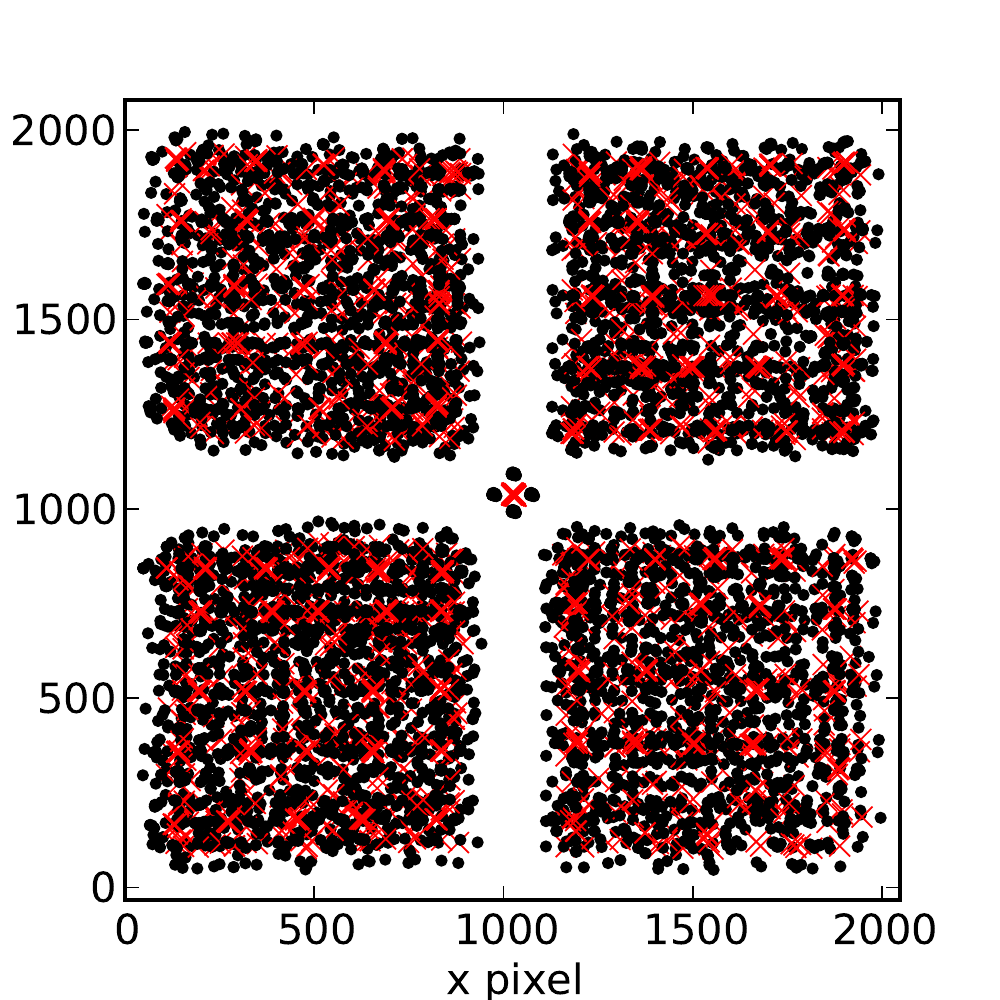}
 \caption{A plot showing the location of unpolarized standard stars used for the combined 2013 -- 2017
 instrument model. The individual spots are indicated by black dots, and the central point by a red
cross.}
 \label{fig:raster_map_plot}
\end{figure}

The instrument model accounts for two separate imperfections of the {\em RoboPol} polarimeter: the spatial distortions of the four spots pattern on the CCD, and the instrumental polarization across the FoV. The functional dependencies used for approximation of the instrument response are given in \cite{King2014}. Here we present updated plots of the instrument model parameters obtained from 11 series of exposures of 7 standard unpolarized stars observed at different positions across the FoV in 2013 -- 2017. The locations of these standard stars on the CCD are shown in \ref{fig:raster_map_plot}.

\begin{figure*}
\centering
\includegraphics[width=0.75\textwidth]{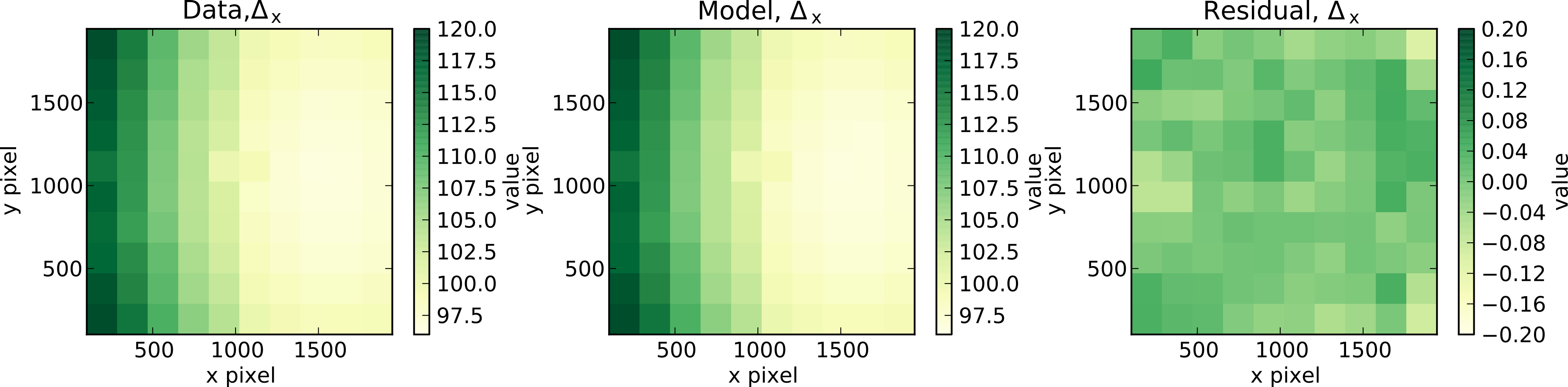}
\caption{The data (left), best-fit model (centre), and residuals (right) for the quantity $\Delta_x$
in the instrument spatial pattern model. Note the change in colour scale for the residual plot.}
 \label{fig:spatial_model:delx}
\end{figure*}

\begin{figure*}
\includegraphics[width=0.75\textwidth]{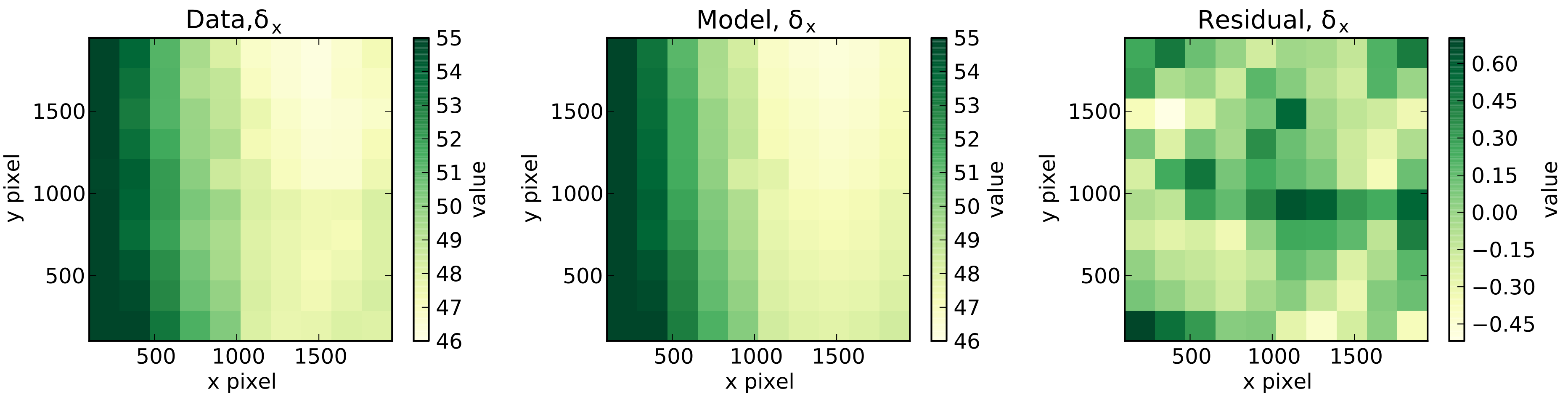}
\caption{The data (left), best-fit model (centre), and residuals (right) for the quantity $\delta_x$
in the instrument spatial pattern model. Note the change in colour scale for the residual plot.}
 \label{fig:spatial_model:dx}
\end{figure*}

\begin{figure*}
\includegraphics[width=0.75\textwidth]{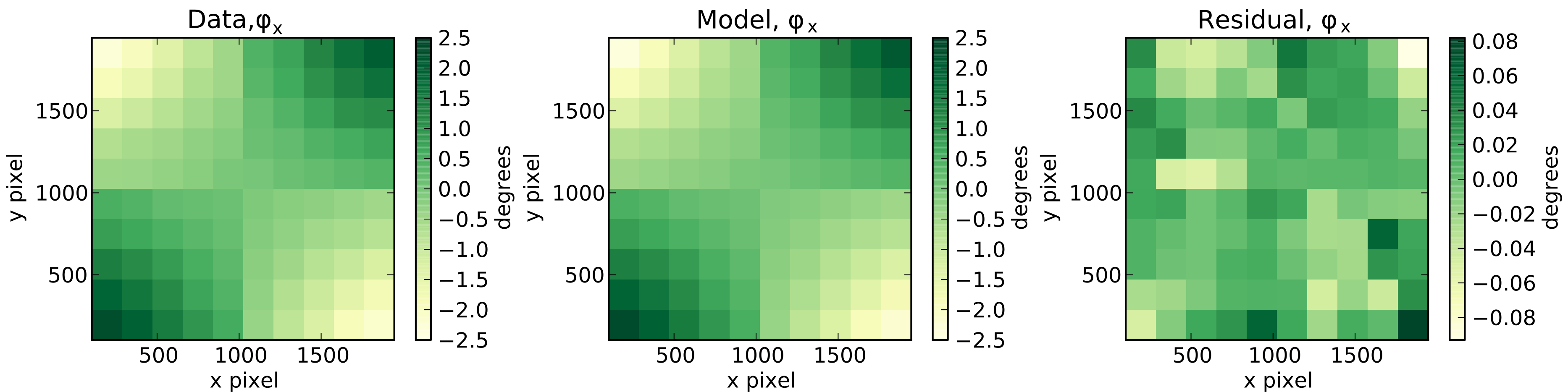}
\caption{The data (left), best-fit model (centre), and residuals (right) for the quantity $\phi_x$
in the instrument spatial pattern model. Note the change in colour scale for the residual plot.}
 \label{fig:spatial_model:phix}
\end{figure*}

Figures \ref{fig:spatial_model:delx} - \ref{fig:spatial_model:phix} demonstrate variation of parameters $\Delta_x(x,y)$, $\delta_x(x,y)$ and $\phi_x(x,y)$ of the spatial model depicted in Fig.~\ref{fig:spot_diagr} across the FoV. The y version of these parameters looks similar. The corresponding plots were omitted for brevity. 

\begin{figure*}
 \centering
\includegraphics[width=0.6\textwidth]{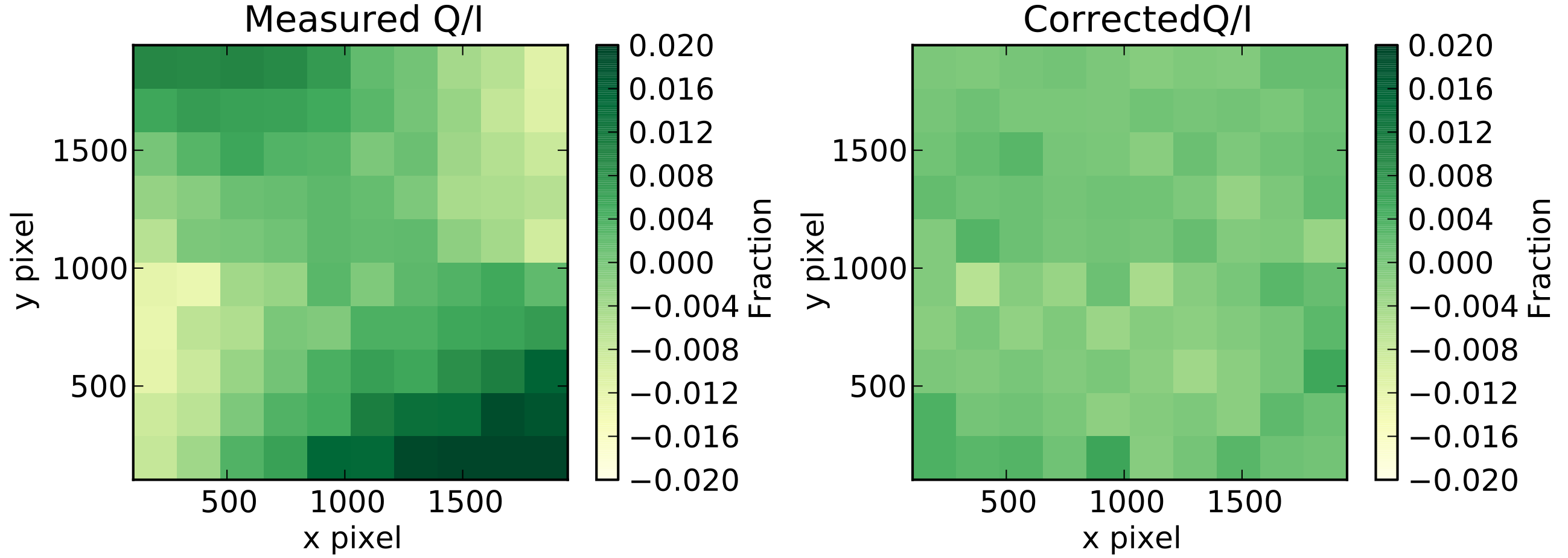}
\caption{The uncorrected (left) and corrected (right) relative Stokes Q/I parameter, which
corresponds to before and after applying the instrument intensity model to the data, respectively.}
 \label{fig:intensity_model:q}
\end{figure*}

\begin{figure*}
 \centering
\includegraphics[width=0.6\textwidth]{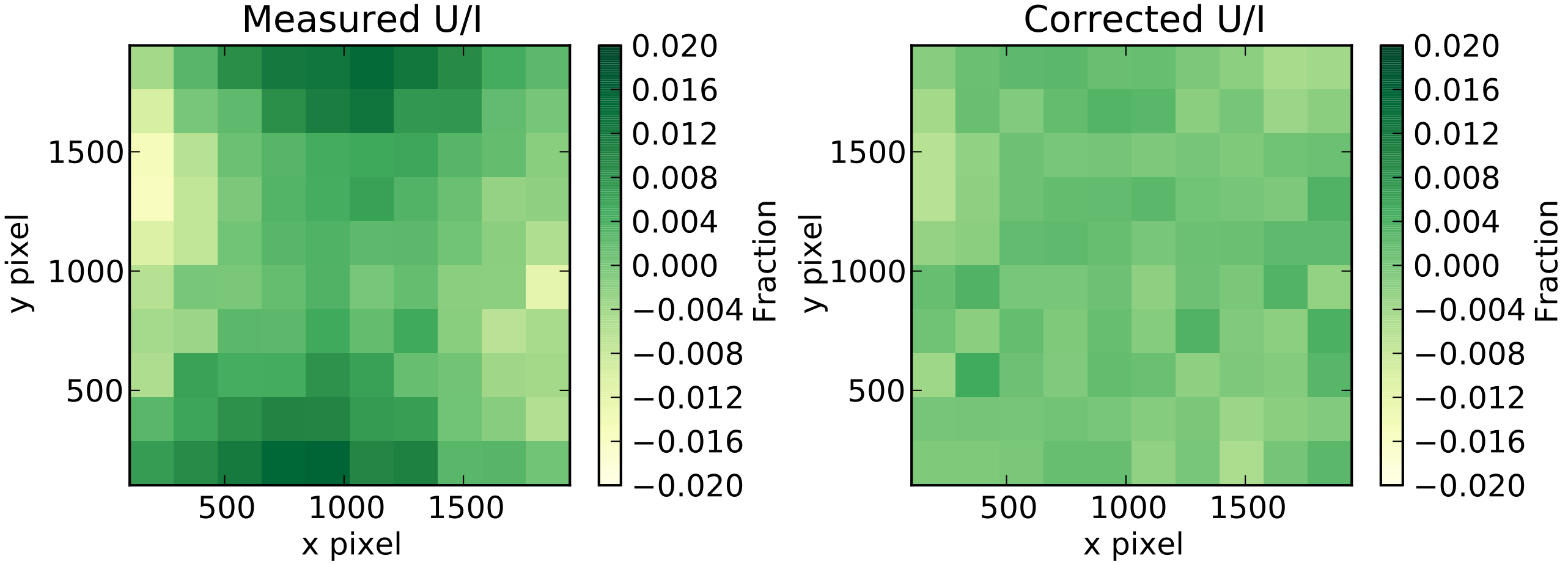}
\caption{The uncorrected (left) and corrected (right) relative Stokes U/I parameter, which
corresponds to before and after applying the instrument intensity model to the data, respectively.}
 \label{fig:intensity_model:u}
\end{figure*}

\begin{figure}
\centering
\includegraphics[width=0.38\textwidth]{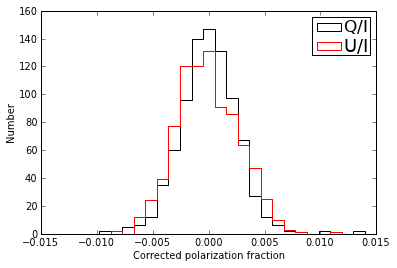}
 \caption{Residuals of relative Stokes parameters after the instrument model fit.}
 \label{fig:qu_res_hist}
\end{figure}

Figures \ref{fig:intensity_model:q} and \ref{fig:intensity_model:u} demonstrate effectiveness of the instrument intensity model used to correct the measured spot intensities for systematic variation of the instrumental polarization across the FoV. The distribution of the residual Q/I and U/I Stokes parameters of the unpolarized standards measured in the field after the instrumental polarization correction is shown in Fig.~\ref{fig:qu_res_hist}. For the detailed description of the instrument model correction we forward readers to \cite{King2014}. 

%%%%%%%%%%%%%%%%%%%%%%%%%%%%%%%%%%%%%%%%%%%%%%%%%%

% Don't change these lines
\bsp	% typesetting comment
\label{lastpage}
\end{document}